# Artificial intelligence-based locoregional markers of brain peritumoral microenvironment

Zahra Riahi Samani[*,1,3], Drew Parker[1,3], Hamed Akbari[2,3], Spyridon Bakas[2,3,5], Ronald L. Wolf[3], Steven Brem[4] and Ragini Verma[1,3]

1. Diffusion & Connectomics In Precision Healthcare Research (DiCIPHR) Lab, University of Pennsylvania, Philadelphia, PA 19104, USA; zari@upenn.edu, william.parker@pennmedicine.upenn.edu; ragini@pennmedicine.upenn.edu
2. Center for Biomedical Image Computing and Analytics (CBICA), University of Pennsylvania, Philadelphia, PA 19104, USA; akbariha@upenn.edu; spyridon.bakas@pennmedicine.upenn.edu
3. Department of Radiology, Perelman School of Medicine, University of Pennsylvania, Philadelphia, Pennsylvania, PA 19104, USA; ronald.wolf@pennmedicine.upenn.edu
4. Department of Neurosurgery, Perelman School of Medicine, University of Pennsylvania, Philadelphia, Pennsylvania, PA 19104, USA; steven.brem@pennmedicine.upenn.edu
5. Department of Pathology & Laboratory Medicine, Perelman School of Medicine, University of Pennsylvania, Philadelphia, PA 19104, USA.

Corresponding Author: Zahra Riahi Samani, email: zari@upenn.edu

**Abstract**
In malignant primary brain tumors, cancer cells infiltrate into the peritumoral brain structures which results in inevitable recurrence. Quantitative assessment of infiltrative heterogeneity in the peritumoral region, the area where biopsy or resection can be hazardous, is important for clinical decision making. Previous work on characterizing the infiltrative heterogeneity in the peritumoral region used various imaging modalities, but information of extracellular free water movement restriction has been limitedly explored. Here, we derive a unique set of Artificial Intelligence (AI)-based markers capturing the heterogeneity of tumor infiltration, by characterizing free water movement restriction in the peritumoral region using Diffusion Tensor Imaging (DTI)-based free water volume fraction maps. A novel voxel-wise deep learning-based peritumoral microenvironment index (PMI) is first extracted by leveraging the widely different water diffusivity properties of glioblastomas and brain metastases as regions with and without infiltrations in the peritumoral tissue. Descriptive characteristics of locoregional hubs of uniformly high PMI values are extracted as AI-based markers to capture distinct aspects of infiltrative heterogeneity. The proposed markers are applied to two clinical use cases on an independent population of 275 adult-type diffuse gliomas (CNS WHO grade 4), analyzing the duration of survival among Isocitrate-Dehydrogenase 1 (*IDH1*)-wildtypes and the differences with *IDH1*-mutants. Our findings provide a panel of markers as surrogates of infiltration that captures unique insight about underlying biology of peritumoral microstructural heterogeneity, establishing them as biomarkers of prognosis pertaining to survival and molecular stratification, with potential applicability in clinical decision making.

**Introduction**
The tumor burden of diffuse gliomas extends beyond the radiographically visible border of tumor margin [1,2,3]. However, current clinical practice considers the tumor core (T1 contrast enhancing boundary) as the primary target of treatment [4,5]; therefore, infiltrative cancer in the peritumoral region (peritumoral T2 hyperintense tissue) remains untreated which leads to inevitable recurrence. Obtaining tissue biopsies can be hazardous in the peritumoral region due to the possibility of functional deficit [4,5]. As a result, characterization of the infiltrative heterogeneity in the peritumoral region is a critical need to inform clinical decision making. Previous attempts to characterize tumor infiltration applied various imaging modalities using manually delineated or heuristic-based labeling of infiltrative tissue [6,7,8,9,10,11,12], but approaches using Diffusion Tensor Imaging (DTI) have been limited to clinically used measures [7,8,9,10].

DTI is the premier Magnetic Resonance Imaging (MRI) modality that provides insight into tissue microstructure by measuring water diffusivity. In particular, the free water volume fraction, a voxel-wise measure of the amount of extracellular water [13], is able to capture differences between infiltrative and vasogenic peritumoral regions by exploiting the variation in water movement restriction [14,15]. The overarching goal of this paper is to leverage this unique information of DTI and derive artificial intelligence (AI)-based markers to capture infiltrative heterogeneity using the extracellular water-based voxel-wise characterization of tissue in the peritumoral region.

By leveraging the differences in water diffusivity properties in the peritumoral region of brain metastases and glioblastomas (Central Nervous System (CNS) World Health Organization (WHO) grade 4, Isocitrate-Dehydrogenase 1 (*IDH1*)-wildtype); consisting of purely vasogenic versus infiltrative edema respectively, we train a deep learning model to derive a novel voxel-wise peritumoral microenvironment index (PMI) without using any manual interaction. The PMI exploits characterization of water movement restriction in the voxels with and without infiltration and hence captures the infiltrative heterogeneity in the peritumoral region. Locoregional hubs of uniformly high PMI values are extracted as regions with high infiltration, and their descriptive characteristics are calculated as AI-based markers of infiltrative heterogeneity, including number and size of hubs and the differences in their shape, direction, and spatial location.

The proposed AI-based markers are applied to two clinical use cases on an independent population of 275 adult-type diffuse gliomas (CNS WHO grade 4) to demonstrate their potential in capturing distinct locoregional aspects of infiltrative heterogeneity beyond standard diffusion measures: (1) analysis of the duration of survival among 264 *IDH1*-wildtype glioblastomas, hypothesizing that higher infiltration of cancer cells, as measured by the proposed markers, is associated with shorter survival; and (2) differences across 275 patients with varying *IDH1* mutation status (astrocytoma *IDH1*-mutant versus glioblastoma *IDH1*-wildtype), notably, the *IDH1*-wildtype tumors, when compared with *IDH1* mutants, have poorer prognosis, and higher peritumoral infiltration [16, 17].

**Results**

The proposed voxel-wise PMI map was generated for independent patients, following the training of a Convolutional Neural Network (CNN) on automatically extracted patches from the peritumoral region of free water volume fraction maps of brain metastases and glioblastomas (CNS WHO grade 4, *IDH1*-wildtype). Figure 1 provides our pipeline and further methodological details can be found in the 'Materials and Methods' section. Following the generation of the PMI map, we identified locoregional hubs of uniformly high PMI values. Descriptive characteristics of these locoregional hubs were extracted as AI-based markers for each patient. These characteristics comprised the number and size of the hubs, as well as the differences in their i) shape (quantified by their individual anisotropic property), ii) directionality, and iii) spatial location (see Figure 2 for a schematic view and 'Materials and Methods' for details). The proposed AI-based markers were evaluated on two clinical use cases to investigate whether the differences in infiltrative heterogeneity, as captured by the proposed descriptive characteristics, are able to capture differences overall survival and *IDH1* mutation status of adult-type diffuse glioma patients.

**Use case 1: AI-based locoregional markers to characterize survival differences in *IDH1*-wildtype grade 4 gliomas**

The first use case included a population of 264 *IDH1*-wildtype grade 4 glioma patients in the survival range of 0.43 to 76.9 months. The goal was to determine whether the proposed descriptive characteristics of PMI locoregional hubs contain information pertaining to patient survival. Statistical analysis across the population, divided into short and long-survival groups at the median, demonstrated significant differences in the proposed descriptive characteristics, using t-test and Bonferroni correction for multiple comparison. The short-survival group had significantly lower number of locoregional hubs ($t_{number}$=2.54, $p_{number}$=0.01) and shape and directional heterogeneity were significantly higher in the short-survival group ($t_{shape}$=3.89, $p_{shape}$<0.001, $t_{directional}$=2.79, $p_{directional}$=0.005). Linear regression analysis of the descriptive characteristics using sex and age as covariates revealed significant differences between short and long-survival groups (Figure 3(a)). The short-survival group had significantly lower number of locoregional hubs ($t_{number}$=2.654, $p_{number}$=0.008). Shape and directional heterogeneity were significantly higher in the short-survival group ($t_{shape}$=3.74, $p_{shape}$<0.001, $t_{directional}$=2.449, $p_{directional}$=0.015). Size and spatial heterogeneity were not found to be significantly different. Examples of PMI maps (3D and overlaid with structural MRI), along with fractional anisotropy (FA), mean diffusivity (MD) and T2 weighted fluid attenuated inversion recovery (T2-FLAIR) images, for a short and a long-survival patient are presented in Figure 3(b). Kaplan-Meier estimates [6] of the two clusters of patients generated downstream of the integration of these descriptive characteristics by K-means clustering are provided in Figure 3(c). A logrank test showed a significant difference between survival distributions of the two clusters (t=19.9, $p<10^{-5}$), and the Cox hazard ratio was 1.82 (95% confidence interval: 1.39, 2.37; $p$ <0.005). Group differences among the two clusters, named as high and low-PMI, were similar to short and long-survival groups, respectively (Figure 3(c) for details and further information in S.1). The cross-correlations between the descriptive characteristics showed the highest correlations between shape heterogeneity and directional heterogeneity, as well as the number and size of the locoregional hubs. However, none of the characteristics described more than 43% of variation in others, as measured by coefficient of determination ($R^2$, details in S.2).

**Use case 2: AI-based locoregional markers to characterize grade 4 gliomas with different *IDH1* mutation status**

The second use case was a population of 275 CNS WHO grade 4 adult-type diffuse gliomas with different *IDH1* mutation status (i.e., *IDH1*-wildtype glioblastomas vs *IDH1*-mutant astrocytomas) to investigate whether the descriptive characteristics of the PMI locoregional hubs can characterize mutation status. Statistical analysis across the population, using t-test and Bonferroni correction for multiple comparison, demonstrated significant differences in the proposed descriptive characteristics between *IDH1*-mutants and *IDH1*-wildtypes. The *IDH1*-wildtype group had significantly lower number of locoregional hubs ($t_{number}$=3.01, $p_{number}$=0.01) and shape heterogeneity and directional heterogeneity were significantly higher in *IDH1*-wildtypes ($t_{shape}$=4.00, $p_{shape}$=0.002, $t_{directional}$=8.7, $p_{directional}<10^{-8}$). Linear regression analysis of the descriptive characteristics using sex and age as covariates revealed significant differences between the two groups (Figure 4(a)). *IDH1*-wildtypes had significantly lower number of locoregional hubs, when compared with *IDH1*-mutants ($t_{number}$=2.861, $p_{number}$=0.005). Shape heterogeneity and directional heterogeneity were significantly higher in *IDH1*-wildtypes comparing to *IDH1*-mutants ($t_{shape}$=2.407, $p_{shape}$=0.017, $t_{directional}$=2.380, $p_{directional}$=0.018). Size and spatial heterogeneity were not found to be significantly different. Examples of PMI maps (3D and overlaid with structural MRI), along with the corresponding FA, MD, and T2-FLAIR images for two patients (*IDH1*-mutant astrocytoma and *IDH1*-wildtype glioblastoma) are presented in Figure 4(b). Descriptive characteristics of the complete population of *IDH1*-mutants versus *IDH1*-wildtype long-survival group versus *IDH1*-wildtype short-survival group are provided in S.3.

**Discussion**

We introduced fully automated, novel AI-based markers of the peritumoral microenvironment (PME) of adult-type diffuse gliomas using hitherto unexplored information of water restriction extracted from DTI. The markers were based on descriptive characteristics derived from locoregional hubs of the PMI map that captured a unique aspect of the infiltrative heterogeneity from the extracellular water properties of the peritumoral regions of metastases and glioblastomas. Our findings identified a panel of markers as surrogates of infiltration, that were established as biomarkers of prognosis based on survival and *IDH1* mutation status.

Previous methods in characterizing the infiltrative heterogeneity utilized various imaging modalities using manual or heuristic-based labeling of infiltrative tissue [6, 7, 8, 9, 10, 11, 12]. DTI is the premier modality to characterize water movement, however, the extracellular free water movement differences have not been explored for peritumoral region. Therefore, the proposed AI-based markers presented here quantified the heterogeneity in the PME which could complement and further strengthen other diffusion-based measures and MRI parameters to aid in personalized treatment planning toward optimizing patient outcome.

Peritumoral edema, a common phenomenon in many malignant tumors, happens due to excess accumulation of fluid in the brain parenchyma resulting from infiltrating tumor cells as well as biological responses to the permeability of the spatially adjacent tumor cells. The spatial distribution and level of water dynamics pattern in edema has not been well defined yet. Using a CNN model trained by characteristics of water diffusivity in glioblastomas versus brain tumor metastases, we were able to capture peritumoral heterogeneity in adult-type diffuse gliomas, without using any manual labeling, to capture surrogates of infiltration and learn how brain parenchyma and immunes system responded to different malignancies.

Brain peritumoral tissue is under the influence of regionally heterogeneous microenvironment. Accurately quantifying this heterogeneity can be crucial for understanding tumor progression. The proposed markers described regionally distinct water movement hubs containing voxels binding to locoregional brain tissue microstructures and we demonstrated their utility in two clinical use cases.

It is well-proved that higher infiltration of tumor cells is associated with poorer prognosis [6, 7, 18, 19], which is consistent with the PMI maps of patients with short and long survival, that showed higher PMI values in the voxels of the short-survival patients. This is compatible with previous literature reporting differences in the levels of FA and MD in altered infiltrated tissues which might be related to higher cellularity and lower water content [7, 8, 20, 21]. Analysis of the descriptive characteristics of PMI locoregional hubs between poor prognosis (short-survivors and high-PMI cluster) and good prognosis (long-survivors and low-PMI cluster), showed a significantly lower number of locoregional hubs in the poor prognosis group which implies larger hubs corresponding to a more infiltrative and aggressive type of tumor. A significantly higher value of shape heterogeneity in the poor prognosis group could reflect the fact that in adult-type diffuse gliomas, cancer cells mainly infiltrate along the white matter [22, 23, 24]. Similarly, higher value of directional heterogeneity for poor prognosis patients could reflect the fact that cancer spreads in different directions comparing to good prognosis group. This is consistent with previous literature showing that glioma cells infiltrate brain tissue by at least two topographic paths, including perivascular invasion along the vascular system or infiltration along the extracellular matrix, nerve, and astrocytic tracts [25].

*IDH1*-wildtype glioblastoma has a poorer prognosis than the *IDH1*-mutant astrocytoma [16, 17, 18] which is consistent with higher PMI values in *IDH1*-wildtypes. The significantly different descriptive characteristics between *IDH1*-mutants and *IDH1*-wildtypes suggests them as powerful markers to discriminate mutation status, even prior to biopsy. *IDH1*-wildtypes had a lower number of locoregional hubs, which could reflect bigger size of adjacent tumoral cells compared to *IDH1*-mutants. This might represent extracellular vesicles which are involved in the rich network of intercellular connections and develop pathologic cascade leading to neurological diseases [26]. Similarly, higher shape, as well as directional heterogeneity in *IDH1*-wildtypes, could demonstrate that in *IDH1*-wildtypes, cancer spreads more throughout the surrounding brain tissue, reflecting poorer prognosis which might result in higher tumor cellularity and infiltration, causing FA and MD changes [27, 28].

The locoregional hubs-based markers introduced here capture the heterogeneity in masses of adjacent voxels with high infiltration which could correspond to a connected set of glioma cells in the peritumoral region. Multicellular networks with filamentous microtubes are found to connect tumoral cells to a network in the peritumoral space in animal models [25, 29]. These cellular networks allow rapid progression and are likely related to treatment resistance [25, 29]. Therefore, biologically applicable markers can be identified in highly connected tumor cells that provide insights into the tumor microenvironment and offer translation to the clinic [25, 29].

From the descriptive characteristics of locoregional hubs, that could be interpreted as signatures of connected tumoral cells, size and number of hubs quantitated modularity, a measure of the structure of networks, and three different aspects of heterogeneity were captured in shape, directionality, and spatial location of hubs. Significantly lower number of hubs in the poor prognosis group demonstrated higher modularity; and significantly higher values of shape and directional heterogeneity suggest connected tumoral cells with higher shape and directional heterogeneity as potentially important biomarkers that could be further explored as key targets for treatment planning and patient selection for clinical trials, including novel immunotherapy and anti-invasion therapy [25, 29, 30].

The data in this study was collected from a single institution. This limitation was addressed by utilizing different acquisition settings and use of an independent cohort for testing, which provides confidence that these results will generalize well to other populations. Our results are strengthened by the fact that they are fully automatic and do not require any manual intervention or reference labels and are easily translatable to the clinic. Thus, although the proposed markers without a corresponding resected specimen cannot pathologically prove the infiltrative heterogeneity of the peritumoral region, they can be used as biomarkers for prognosis.

In the future, the proposed markers could be integrated with other imaging modalities to provide biologically relevant characterization of the PME to quantitate invasion and microenvironmental heterogeneity for effective cancer therapy [31, 32, 33]. Further studies using the markers described here could be useful to elucidate biological processes linked to the *IDH1* mutation status and better understand multiple sources of heterogeneity in adult-type diffuse gliomas which has major clinical implications [34, 35, 36].

**Materials and Methods**
**Datasets**
This study was approved by the institutional review board of University of Pennsylvania. Informed consent was obtained from all participants or their legally authorized representative. All methods were carried out in accordance with relevant guidelines and regulations. The population was identified based on retrospective review of the electronic medical record of patients diagnosed with adult-type diffuse glioma (CNS WHO grade 4). Study inclusion criteria included histopathological tissue diagnosis of adult-type diffuse glioma (CNS WHO grade 4) or brain metastasis and preoperative MRI at time of diagnosis including structural and diffusion MRI. For the *IDH1* mutation study, sufficient tumor tissue collected at time of surgery was required. We identified 381 patients with adult-type diffuse gliomas (CNS WHO grade 4) and 50 patients with brain metastases, and they were randomly divided into three datasets. **The training dataset** was used to train the CNN and included 106 patients with brain tumors, 66 *IDH1*-wildtype glioblastomas, aged 23–83 years (mean: 60.5, standard deviation (SD): 11.8), and in a male: female proportion of 37:29; and 40 metastases, aged 29–87 years (mean: 62.12, SD: 12.6), and in a male: female proportion of 18:22. **The locoregional hub dataset** was used to find the optimal threshold to make locoregional hubs. This data set was independent from the training dataset and consisted of 30 patients with brain tumors, 20 *IDH1*-wildtype glioblastomas and 10 brain metastases, aged 42-84 years (mean: 64.3, SD: 10.5), and in a male: female proportion of 16:14. **The test data set** was used for two different use cases, comprising survival among *IDH1*-wildtype glioblastomas and differences among *IDH1*-mutation status. **The survival use case** included 264 *IDH1*-wildtype glioblastomas in the survival range of 0.43 to 76.9 month, aged 21-88 years (mean: 63.6, SD: 11.4), and in a male: female proportion of 160:104. **The mutation use case** consisted of 275 CNS WHO grade 4 adult-type diffuse gliomas, 264 patients with *IDH1*-wildtype glioblastoma as in the survival use case and 11 *IDH1*-mutant astrocytomas, aged 21-88 years (mean: 62.5, SD: 12.5), and in a male: female proportion of 165:110. The acquisition parameters for all datasets were based on single-shell diffusion data on two Siemens 3T scanners, Verio or TrioTim, with TR/TE in range of 4200-7400 ms/83-88 ms, with 1 unweighted volume and either 30 or 12-direction diffusion-weighted volumes, at a b-value of 1000 s/mm$^2$. The acquisition was repeated between 1 and 6 times for improved signal to noise ratio (SNR). The spatial resolution was 1.7x1.7x3 mm.

**Creation of free water map and masks of tumor and edema**
All diffusion data was pre-processed using local Principal Component Analysis (PCA) denoising [37], eddy current and motion correction using FSL EDDY [38], and skull-stripping with Brain Extraction Tool (BET) [39]. Fractional anisotropy (FA) and mean diffusivity (MD) maps were calculated after DTI fitting with DIPY using weighted least squares [40]. Structural scans (post-contrast T1, T2, and T2-FLAIR) were registered to the pre-contrast T1 with rigid registration in Advanced Normalization Tools (ANTs) [41, 42], and then registered to the FA image with a nonlinear registration to account for Echo Planar Imaging (EPI) distortions in the diffusion data in ANTs. Masks of the tumor and edema for each patient were created from the registered structural images using Deep-Medic [43]. We used Freewater EstimatoR using Interpolated Initialization (FERNET) [13], a free water elimination paradigm designed for single-shell diffusion MRI data using a novel interpolated initialization approach, to estimate the free water compartment in single-shell diffusion MRI data. FERNET provides the user with a free water volume fraction map that were resampled to 2x2x2 mm spatial resolution.

**Creation of the PMI map and locoregional hubs**
The training dataset was used to train the deep learning model to extract the PMI map. We automatically extracted a set of (16×16) voxel patches from the peritumoral edema of metastases and *IDH1*-wildtype glioblastomas placed at random locations and random directions (within axial, sagittal and coronal planes) using random seed generators. This was the largest patch that could be fit into edema without overlapping into the main tumor mass. Patches in the peritumoral edema of metastases and glioblastomas were labeled as high free water and low free water, respectively. The convolutional neural network consisted of 6 convolutional layers followed by a max-pooling and a fully connected layer. A softmax layer at the end produced a probability value for every input patch that indicated its membership to each class, either high free water or low free water [44]. Data augmentation was done on the patches by shifting them in random directions, letting a maximum of 20% overlap with the healthy brain. The hyper-parameters of CNN were weight decay $5\times10^{-5}$, momentum 0.9, initial learning rate $10^{-4}$.

To obtain the PMI map for an unseen patient, we placed patches centered at each voxel within the sagittal, axial, and coronal directions. The PMI value for each voxel was then calculated by averaging among the result of the CNN for the patches.

For getting locoregional hubs, we extracted connected components (CCs) of high PMI values [45]. We made CCs for patients in the locoregional hub dataset in different thresholds in range of [0 1] by steps of 0.1. Connected components with diameters less than 2 voxels were not taken into consideration. The threshold which provided the most significant differences in the descriptive characteristics between 30 samples of metastases and glioblastomas was chosen (threshold=0.9) and used for the mutation and the survival use cases.

**Extracting the descriptive characteristics**
The number and size of the locoregional hubs were normalized based on the number of voxels with PMI values greater than the threshold. The size was averaged among the hubs of each patient. Shape heterogeneity, directional heterogeneity, and spatial heterogeneity were computed as follows: *Shape heterogeneity:* standard deviations of voxels coordinate in each hub along three (axial, sagittal, coronal) directions were calculated. Shape heterogeneity was subsequently defined as the normalized difference between the maximum and second-largest standard deviation and averaged among hubs of each patient. *Directional heterogeneity*: a three-dimensional feature vector was calculated for each hub, consisting of standard deviations of voxels coordinate along the three (axial, sagittal, coronal) directions. Hausdorff distance [46] was applied on the pair-wise cosine distance among hubs to compute directional heterogeneity.

*Spatial heterogeneity*: the center of gravity for each hub and the average Euclidean distance among them was calculated. This distance was further normalized by the diameter, or maximum possible distance between any two points, of the edema region to get spatial heterogeneity.

**Analysis of survival and mutation**
The PMI maps were generated for the patients in the survival and mutation use cases. 8 patients were removed from both use cases as they had small edema size (less than two voxels around tumor). Next, the locoregional hubs of high PMI values were generated. 9 patients from the survival use case and 10 patients from the mutation use case were removed as they had less than 2 hubs which was the minimum number needed to compute directional and shape heterogeneity. To investigate the significance of the differences: (i) two sample two tailed t-test was done for each of the descriptive characteristics, followed by Bonferroni correction for multiple comparison; (ii) linear regression was performed on each of the descriptive characteristics with sex and age as covariates. Kaplan-Meier curves were fit using nonparametric Turnbull estimator [47]. Clustering was done for different number of clusters in the range of 1 to 9 and the one that maximized the Calinski-Harabasz index [48] was chosen.

**Acknowledgments**
Research reported in this publication was supported by the National Institutes of Health (NIH) Grant R01NS096606 (PI: Ragini Verma), National Cancer Institute (NCI) of the National Institutes of Health (NIH) under award numbers: NCI: U01CA242871, NCI: U24CA189523, NINDS: R01NS042645. The content of this publication is solely the responsibility of the authors and does not represent the official views of the NIH.

**Author contributions**
**Conceptualization:** Z.R.S., R.V.; **Methodology:** Z.R.S.; **Experiments:** Z.R.S.; **Data processing:** D.P.; **Analysis:** Z.R.S., D.P., H.A.; **Interpretation:** Z.R.S., D.P., H.A., Sp.B., R.L.W., S.B., R.V.; **Writing and editing:** Z.R.S., D.P., H.A., Sp.B., R.L.W., S.B., R.V.; **Supervision:** S.B., R.V.

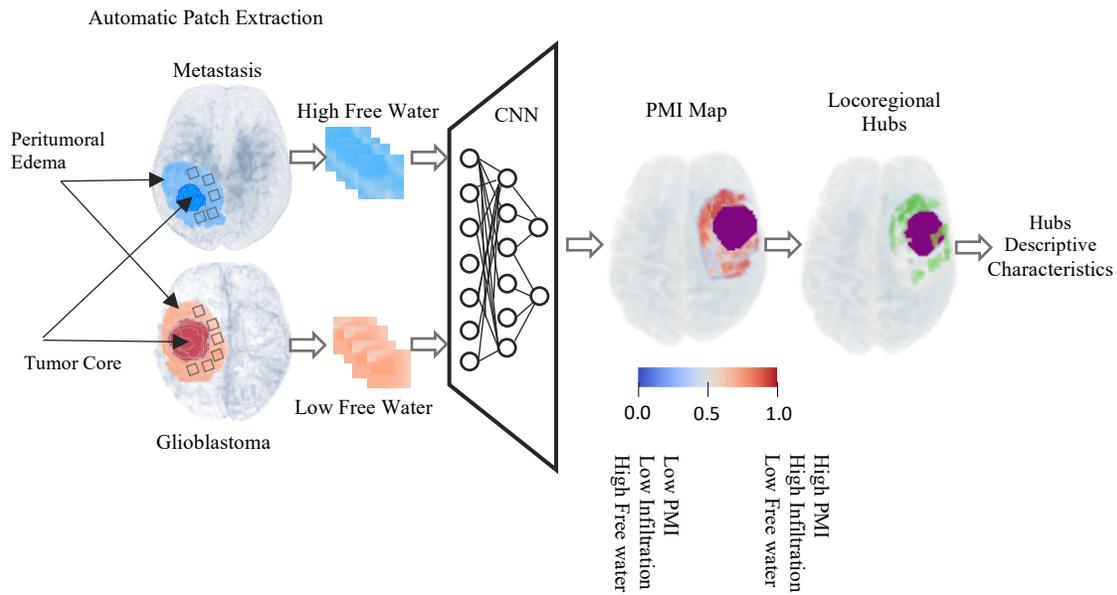

**Figure 1.** The pipeline for the generation of the Peritumoral Microenvironment Index (PMI) map and locoregional hubs. The inputs to the Convolutional Neural Network (CNN) are patches (boxes) extracted from the free water volume fraction map in the peritumoral region from both glioblastomas (red) and metastases (blue) labeled as low free water and high free water which are used to train the CNN. Locoregional hubs are extracted from PMI. Descriptive characteristics of the locoregional hubs are extracted as AI-based markers.

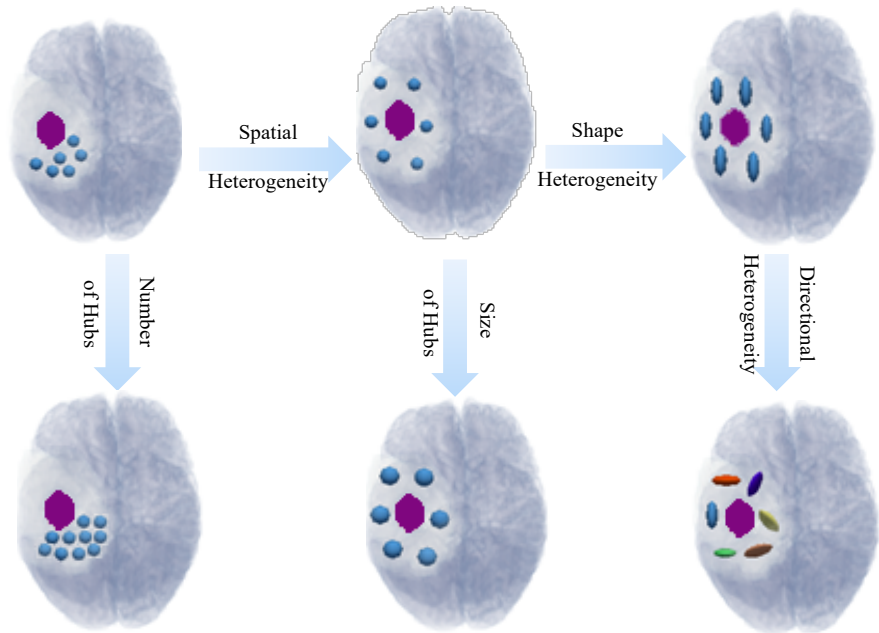

**Figure 2.** A schematic view of locoregional hubs descriptive characteristics: number of hubs, size of hubs, shape heterogeneity, directional heterogeneity, and spatial heterogeneity. The purple circle displays the tumor core and blue hubs are located in the peritumoral edema. Arrows are directed toward increasing the values of the descriptive characteristics. Figure in the top-left is the reference figure for the hubs. (i) Moving from left to right, to the top-middle: the spatial heterogeneity of hubs increases; to the top-right: the shape heterogeneity increases. (ii) Moving from top to the bottom, left: the number of hubs increases; center: size of hubs increases; right: directional heterogeneity increases.

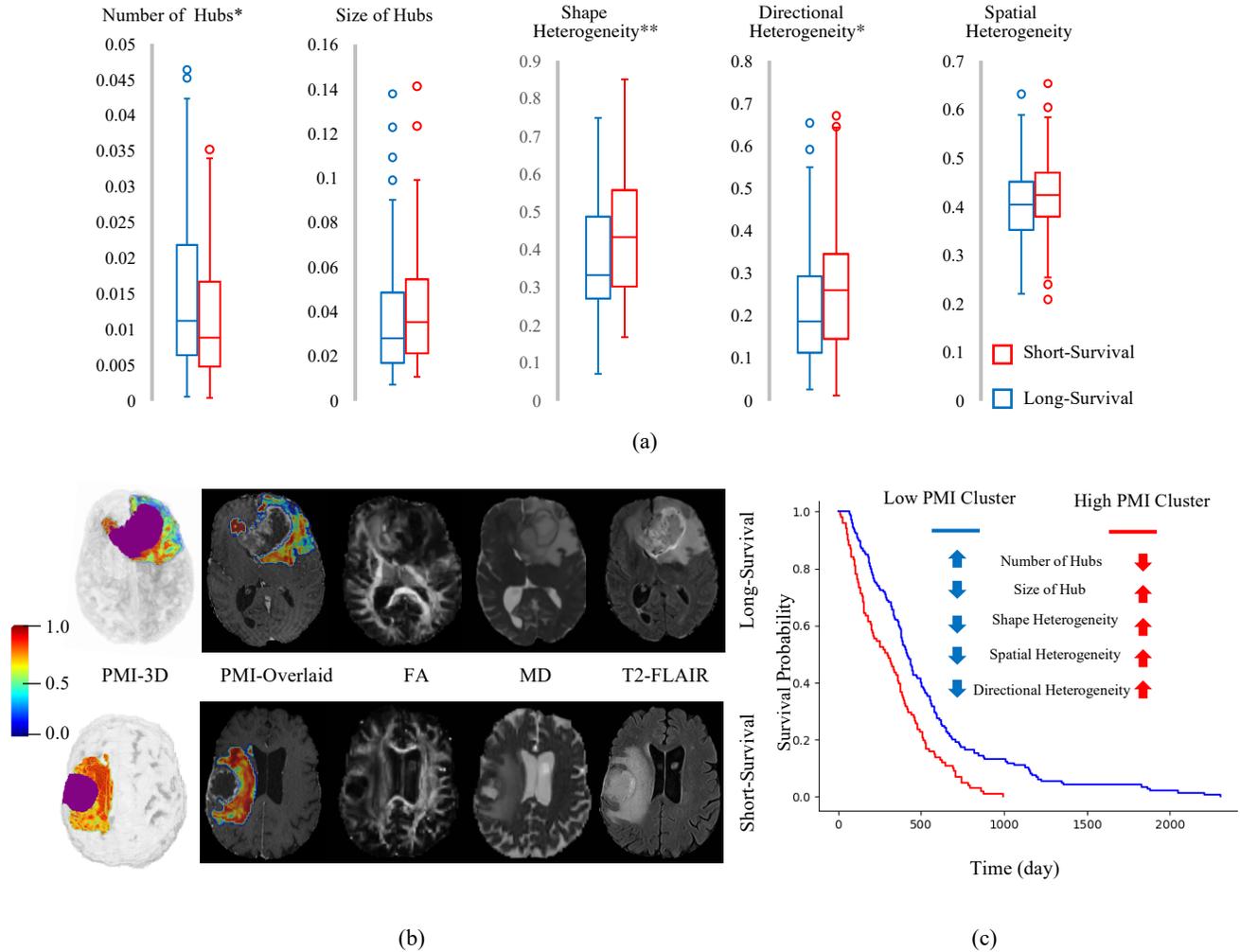

**Figure 3.** The PMI map and AI-based markers for *IDH1*-wildtype grade 4 glioma patients with different duration of survival. (a) Descriptive characteristics of PMI locoregional hubs for long and short survival groups, *p* value <0.05 (*), *p* value<0.005 (**), Linear regression was used with age and sex as covariates. (b) representative samples of the PMI map with FA, MD, and T2-FLAIR images for a long and a short survival patient. (c) Kaplan-Meier estimates of the two clusters generated by K-means clustering.

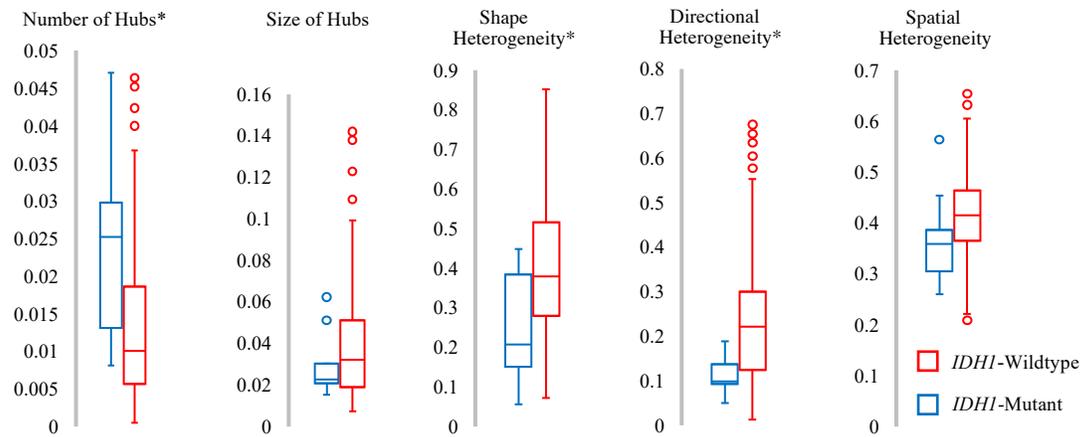

(a)

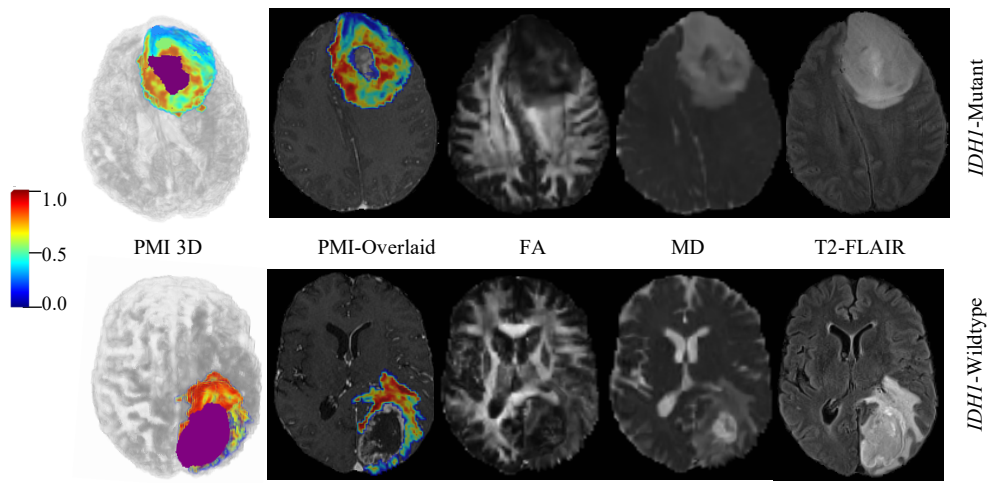

(b)

**Figure 4.** The PMI map and AI-based markers for grade 4 glioma patients with different *IDH1* mutation status. (a) Descriptive characteristics of PMI locoregional hubs for *IDH1*-mutant astrocytoma and *IDH1*-wildtype glioblastoma patients, *p* value <0.05 (*), Linear regression was used with age and sex as covariates. (b) representative samples of the PMI map with FA, MD and T2-FLAIR images for an *IDH1*-mutant astrocytoma and an *IDH1*-wildtype glioblastoma patient.

**Supplementary Materials**

**S.1. Descriptive characteristics of low-PMI and high-PMI clusters**

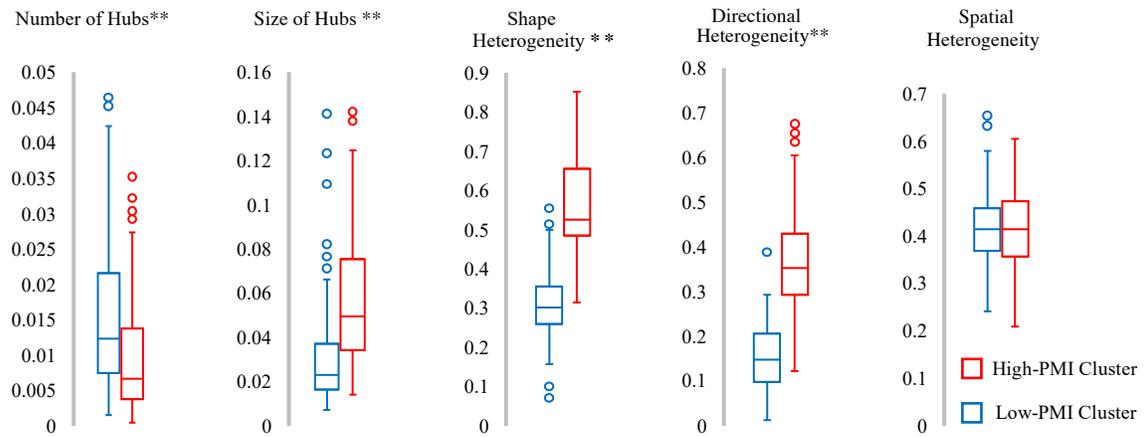

**Figure S.1.** Descriptive characteristics of PMI locoregional hubs for low-PMI and high-PMI clusters (t-test, $p$ value<$10^{-4}$)

## S.2. Cross-Correlations between the proposed descriptive characteristics

|  | Number of Hubs | Size of Hubs | Shape Heterogeneity | Directional Heterogeneity | Spatial Heterogeneity |
|---|---|---|---|---|---|
| Number of Hubs | X | | | | |
| Size of Hubs | -0.34 | X | | | |
| Shape heterogeneity | -0.14 | 0.23 | X | | |
| Directional Heterogeneity | -0.07 | 0.21 | 0.65 | X | |
| Spatial Heterogeneity | 0.17 | -0.24 | 0.02 | 0.01 | X |

## S.3. Descriptive characteristics of *IDH1*-mutant vs *IDH1*-wildtype long-survival vs *IDH1*-wildtype short-survival groups

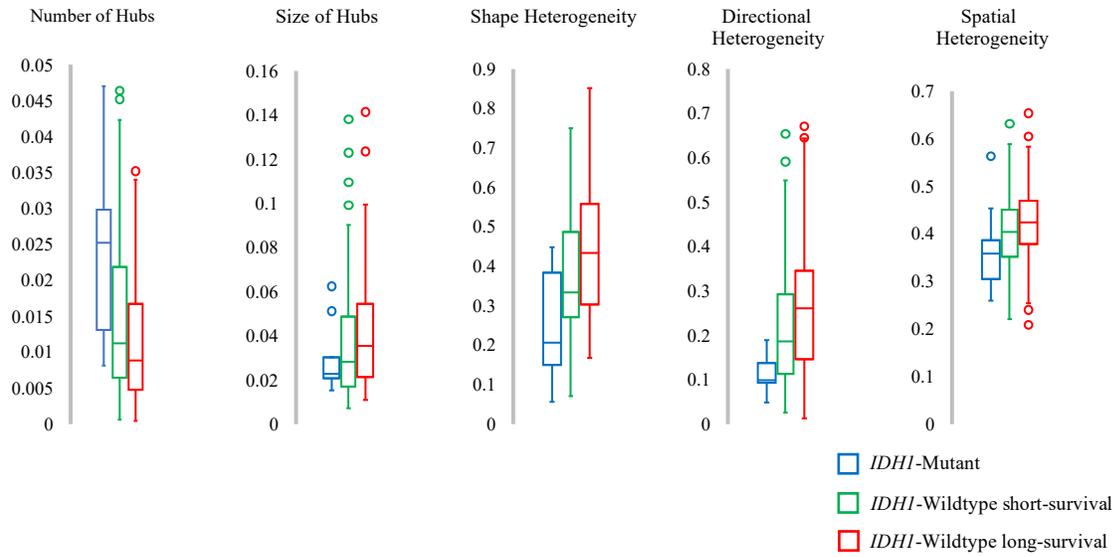

**Figure S.3.** Descriptive characteristics of *IDH1*-mutant vs *IDH1*-wildtype long-survival vs *IDH1*-wildtype short-survival groups